\documentstyle[12pt]{article}

\begin{document}

\begin{titlepage}
\null\vspace{-62pt}

\pagestyle{empty}
\begin{center}
\rightline{hep-ph/0207xxx}

\vspace{1.0truein} {\Large\bf Hard thermal loops with a background
plasma velocity}

\vspace{1in}
D. METAXAS \\
\vskip .4in {\it Physics Department,
National Technical University\\
Zografou Campus, 15780, Athens, Greece\\
email: dkm@central.ntua.gr}\\

\vspace{0.5in}

\vspace{.5in} \centerline{\bf Abstract}

\baselineskip 18pt
\end{center}

I consider the calculation of the two and three-point functions
for QED at finite temperature in the presence of a background
plasma velocity.  The final expressions are consistent with
Lorentz invariance, gauge invariance and current conservation,
pointing to a straightforward generalization of the hard thermal
loop formalism to this physical situation. I also give the
resulting expression for the effective action and identify the
various terms.

\end{titlepage}

\newpage
\pagestyle{plain} \setcounter{page}{1}
\newpage

\section{Introduction}

The calculation of quantities like the photon polarization tensor,
$\Pi_{\mu \nu}$, and the fermion self energy, $\Sigma$, for
arbitrary values of the photon and fermion momentum $(k_0 =
\omega, \, \vec{k})$, at finite temperature is important for the
extraction of many characteristic plasma properties such as the
dielectric response, dispersion relations, and calculation of
plasma radiation \cite{FTFT}.

At finite temperature Lorentz invariance is more involved because
of the presence of the background plasma velocity, $u^{\mu}$.
Usually, the calculations are done for the case of a plasma at
rest, $u^{\mu} = (1, 0, 0, 0)$, in which case Lorentz invariance
is lost because of the existence of a preferred rest frame.
Various quantities can then depend separately on $\omega$ and
$\vec k$.

Applications of finite temperature field theory at realistic
situations, however, would need the extension of these results to
include an arbitrary (determined by the equations of motion)
background plasma velocity. Such would be the case also if one
wants to consider the calculation of various properties of the
quark gluon plasma state formed in relativistic heavy ion
collisions \cite{QGP}.

The inclusion of the plasma velocity effects can be done when one
uses a more general thermal distribution function derived from
kinetic theory \cite{vanweert, weldon}. In the case of fermions,
for example, in the place of the ordinary Fermi-Dirac
distribution, we have

\begin{equation}
n(p) = \frac{1}{\exp{\beta\, |u\cdot p|} +1}
\end{equation}
where $\beta$ is the inverse temperature $T$, and $p$ is the
fermion four momentum.

The covariant expressions for the vacuum polarization and the
fermion self-energy were given in \cite{weldon}, \cite{weldon2}.
Here, besides the two-point functions I will consider the
calculation of the vertex correction in QED, in the hard thermal
loop approximation, and show that the final expressions satisfy
the conditions of gauge invariance that are necessary for the
application of the hard thermal loop resummation.

From the resulting expressions I also calculate the general form
of the hard thermal loop effective action for QED and give the
first order corrections in the nonrelativistic limit,
$u^{\mu}\approx (1,\vec{u})$. In this limit I also give the
expression for the dielectric tensor $\epsilon_{ij}$, which will
now depend also on $u$. The corresponding contribution to the
effective action will be of the form $\epsilon_{ij}\, E_i \, E_j$,
where $\vec{E}$ is the electric field.

There is, also, another term in the resulting effective action,
which, as I will show, turns out to be proportional to
$\vec{u}\cdot(\vec{E}\, \times\,\vec{B})$, where $\vec{B}$ is the
magnetic field. This term is expected to appear in the effective
action of electrodynamics in the case of moving dielectrics
(\cite{diel} and references therein), and was derived there
heuristically using symmetry arguments. Although the physical
situation here is different, the same term appears, and its
coefficient is calculated from the microscopic theory. As shown in
\cite{diel}, the inclusion of this term is important in the
consideration of radiative processes, and it should be significant
in the applications of the present results as well.

In Sec.~2 I will review and outline the calculation of the
two-point functions at finite temperature including the finite
plasma velocity effects, in the formalism appropriate to the hard
thermal loop resummations.
 In Sec.~3 I will give the expression for the three-point
function and show that, this, as well as the previous expressions,
satisfy the constraints imposed by gauge invariance, Lorentz
invariance and current conservation. I also discuss the
contributions to the effective action and give the interpretation
of the resulting effective action in terms of a velocity-dependent
dielectric tensor and the additional term mentioned above.
Finally, in Sec.~4 I will conclude with some remarks about the
possible applications and extensions of these results.

\section{Two-point functions
 with a background plasma velocity}

The fermion propagator at finite temperature can be written as
\begin{equation}
S(x, y) = \int \frac{d^4 p}{(2\pi)^4} \left[ \frac{i(\gamma\cdot
p)}{p^2  + i\epsilon} - 2\pi(\gamma\cdot p) n_p \, \delta (p^2 )
\right] e^{-i p\cdot (x-y) }
\end{equation}
where we have neglected the fermion mass in the high temperature
limit. We use the generalized distribution function
\begin{equation}
n_p = \frac{1}{e^{\beta  |u\cdot p|} +1}
\end{equation}
and the propagator can be evaluated as
\begin{eqnarray}
S(x,y) &=& \int\frac{d^3 p}{(2\pi)^3}\frac{1}{2 E_p} {\Big\{}
 \theta(x^0 -y^0) \left[ \alpha_p (\gamma\cdot p)
 e^{-ip\cdot(x-y)} +\beta_{p'} (\gamma\cdot p') e^{ip'\cdot(x-y)}
 \right]  \nonumber \\
  &-& \theta(y^0 -x^0) \left[ \beta_p (\gamma\cdot p)
 e^{-ip\cdot(x-y)} + \alpha_{p'} (\gamma\cdot p')
 e^{ip'\cdot(x-y)} \right] {\Big\}}
 \label{fp}
 \end{eqnarray}
 where
 \begin{equation}
 \alpha_p = 1-n_p \,,\,\,\,\beta_p =n_p
 \end{equation}
 \begin{equation}
 p=(E_p, \vec{p}) \,,\,\, p'=(E_p, -\vec{p})
 \end{equation}

 The one-fermion loop contribution to the effective action
 is
 \begin{equation}
 \Gamma = \frac{i}{2} \int d^4 x \,d^4 y \,Tr
 \left[ \gamma\cdot A(x) S(x, y) \gamma\cdot A(y) S(y, x)
 \right].
 \end{equation}

We substitute $ A_{\mu} (x) = \int \frac{d^4 k}{(2\pi)^4} e^{i
k\cdot x} A_{\mu}(k)$ and after
 we perform the traces and take
$|\vec{k}|$ small compared to $|\vec{q}|$ in the hard thermal loop
approximation we get
\begin{equation}
\Gamma = \int_{k,k'} \, A_{\mu}(k) A_{\nu}(k')
         (2\pi)^4 \, \delta^{(4)} (k+k') \, \Pi^{\mu\nu} (k)
       \label{eff}
\end{equation}
with the  temperature dependent part of the photon polarization
tensor
\begin{eqnarray}
& & \Pi^{\mu\nu}(k) =-\frac{1}{2}\int \frac{d^3q}{(2\pi)^3}
    \frac{1}{2p_0 2q_0} \nonumber \\
& & {\Big\{}\,\, (n_q-n_p) \frac{8 q_0^2 Q^{\mu}
Q^{\nu}}{p_0-q_0-k_0}
 +(n_{q'}-n_{p'}) \frac{8 q_0^2 Q'^{\mu} Q'^{\nu}}{p_0-q_0+k_0}
    \nonumber \\
& & - 4q_0^2 (Q^{\mu}Q'^{\nu} +Q'^{\mu}Q^{\nu} -2g^{\mu\nu})
    \left( \frac{n_p+n_{q'}}{p_0+q_0-k_0}
         +\frac{n_{p'}+n_q}{p_0+q_0+k_0} \right)\,\,{\Big\}}
\end{eqnarray}
where  $\vec{p}=\vec{q}+\vec{k}$ and we have the following
notations and relations:
\begin{eqnarray}
& & Q^{\mu}=(1, \hat{q}) \,\,\,\,\, Q'^{\mu}=(1, -\hat{q})
  \,\,\,\,\, \hat{q}=\vec{q}/q_0  \nonumber \\
& & p_0 \approx q_0 + \hat{q}\cdot\vec{k}
\end{eqnarray}
We have also dropped the $i\epsilon$'s been interested here in the
real part. Now one can decouple the angular integrals using
relations such as
\begin{eqnarray}
& & n_p -n_q \approx \frac{dn_q}{dq_0}
    \left( \frac{u_0 \, \hat{q}\cdot\vec{k} -\vec{u}\cdot\vec{k}}
                {u_0 - \vec{u}\cdot\hat{q}} \right)
     \nonumber \\
& & \int \frac{d^3 q}{(2\pi)^3} \,\frac{dn_q}{dq_0} f(\hat{q})
    = - \int \frac{d^3 q}{(2\pi)^3} \,\frac{2 n}{q_0} f(\hat{q})
    \nonumber \\
& & \int\frac{d^3 q}{(2\pi)^3} \frac{2 n_q}{q_0}
    = \int d\Omega \frac{T^2}{48\pi}
                   \,\frac{1}{(u_0 -\vec{u}\cdot\hat{q})^2}
\end{eqnarray}
to get the general expressions for the polarization tensor
\begin{equation}
\Pi^{00} = 2m^2\int\frac{d\Omega}{4\pi}
           \,\frac{1}{(u\cdot Q)^3}
           \left( -u_0 + \frac{u\cdot k}
                              {k\cdot Q}
                              \right)
\label{p1}
\end{equation}
\begin{equation}
\Pi^{0i} = 2m^2 \int\frac{d\Omega}{4\pi}
             \,\frac{\hat{q}_i}{(u\cdot Q)^3}
             \left( -u_0 +\frac{u\cdot k}
                               {k\cdot Q}
                              \right)
\label{p2}
\end{equation}
\begin{eqnarray}
\Pi^{ij} = 2m^2\int\frac{d\Omega}{4\pi} &{\Big\{}&
\,\frac{\hat{q}_i \hat{q}_j}{(u\cdot Q)^3}
            \left( -u_0 +\frac{u\cdot k}
                            {k \cdot Q}\right)
\nonumber \\
& & -\frac{1}{2}\, \frac{\hat{q}_i \hat{q}_j +g^{ij}}
                        {(u\cdot Q)^2}
                        {\Big\}}
\label{p3}
\end{eqnarray}
where we have defined the photon thermal mass $m^2 = e^2 T^2/6$.

 These are the general expressions for the polarization tensor.
It is straightforward to verify that they satisfy the condition
imposed by current conservation, $k_{\mu} \, \Pi^{\mu\nu} =0$, and
use it to get the full expression in a Lorentz covariant form
\cite{weldon}. Here I will give the result
 in the limit of nonrelativistic $u$, which will be used in the next
  section, taking $u^{\mu}\approx (1,\vec{u})$,
   and keeping the terms of
first order in $\vec{u}$.

\begin{eqnarray}
\Pi^{00}/(m^2) &=& (-1+L) + \nonumber \\
         &+& (2 k^2 \, F_1) (\vec{u}\cdot\vec{k} )
         \label{p00}
\end{eqnarray}
\begin{eqnarray}
\Pi^{0i}/(m^2) &=& (-1+L)\frac{k_0 k_i}{k^2} + \nonumber \\
               &+& (3k_0 F_2 -1) \, u_i + \nonumber \\
               &+&(3k_0 F_1 -L+1) (\vec{u}\cdot\vec{k}) k_i
          \label{p0i}
\end{eqnarray}
\begin{eqnarray}
\Pi^{ij}/(m^2) &=& (k_0 F_1) k_i k_j + (k_0 F_2) \delta_{ij} +
                                          \nonumber \\
               &+&(3 k_0 G_1 -F_1) (\vec{u}\cdot\vec{k}) k_i k_j +
                                          \nonumber \\
          &+&(3 k_0 G_2 -F_2) (\vec{u}\cdot\vec{k}) \delta_{ij} +
                                      \nonumber \\
         &+& (3k_0 G_2) (u_i k_j +u_j k_i)
        \label{pij}
\end{eqnarray}
where
\begin{equation}
L=\frac{1}{2}\frac{k_0}{|\vec{k}|} \ln\left(\frac{k_0
+|\vec{k}|}{k_0-|\vec{k}|}\right)
\end{equation}
and
\begin{equation}
F_1 =\frac{1}{2k_0 k_i^4}(3k_0^2L-k_i^2L-3k_0^2)
\end{equation}
\begin{equation}
F_2=\frac{1}{2k_0k_i^2}(k_i^2L-k_0^2L +k_0^2)
\end{equation}
\begin{equation}
G_1 =\frac{1}{6k_i^6}(15k_0^2L-9k_i^2L-15k_0^2 +4k_i^2)
\end{equation}
\begin{equation}
G_2=\frac{1}{6k_i^4}(3k_i^2L-3k_0^2L+3k_0^2-2k_i^2)
\end{equation}

The calculation of the fermion self-energy, $\Sigma(k)$, can be
done in the same way, using the photon propagator in the Landau
gauge,
\begin{equation}
G(x, y) =\int \frac{d^4 p}{(2\pi)^4} \left[
         \frac{i}{p^2 +  i\varepsilon}
         + 2 \pi n^B_p \, \delta(p^2) \right]
         e^{-ip\cdot (x-y)}
\end{equation}
where,
\begin{equation}
n^B_p =\frac{1}{e^{\beta |u\cdot p|} -1}
\end{equation}

The calculations are done similarly in coordinate space using the
above two-point function in the form
\begin{eqnarray}
G(x, y)&=&\int \frac{d^3 p}{(2\pi)^3} \frac{1}{2 E_p}
        {\Big\{} \theta(x^0-y^0)\left[ (1+n_p^B) e^{-ip\cdot(x-y)}
        +n_{p'}^B e^{i p' \cdot(x-y)} \right] \nonumber \\
        &-& \theta(y^0-x^0)\left[ n_p^B e^{-ip\cdot(x-y)}
        +(1+n^B_{p'}) e^{ip' \cdot(x-y)} \right] {\Big\}}
\label{bp}
\end{eqnarray}
and the final result for the fermion self-energy is
\begin{equation}
\Sigma(k) = m_f^2 \, \int \frac{d\Omega}{4\pi}
          \frac{1}{(u\cdot Q)^2}
            \frac{\gamma\cdot Q}{k\cdot Q}
\label{fermion}
\end{equation}
with the fermion thermal mass $m_f^2 = e^2 \, T^2 /8$.

This can be evaluated as
\begin{equation}
\Sigma(k) = (\Sigma_1 \, k^{\mu} +\Sigma_2 \, u^{\mu})
           \,\gamma_{\mu}
\label{sfermion}
\end{equation}
where
\begin{equation}
\Sigma_1 = -\frac{m_f^2}{K^2}\,(1-\tilde{L}) \,\,\,\,\,\,\,
 \Sigma_2 = \frac{m_f^2}{K^2}\,(\Omega-\frac{k^2}{\Omega}
                                  \tilde{L})
\end{equation}
with
\begin{equation}
\Omega=(u\cdot k), \,\,\,\,\,\, K=\sqrt{(u\cdot k)^2 -k^2}, \,\,\,
 \,\, \tilde{L}=\frac{\Omega}{2K} \,\ln \left( \frac{\Omega +K}
                      {\Omega -K} \right)
\end{equation}
The poles of the effective fermion propagator
\begin{equation}
S(k)=\frac{1}{\gamma\cdot k - \Sigma(k)}
\end{equation}
give two branches for the dispersion relations in terms of
$\Omega$ and $K$, which are the appropriate relativistic
generalizations of $\omega$ and $|\vec{k}|$. These two branches
$\Omega_{\pm} (K)$ are the solutions of
\begin{equation}
(1-\Sigma_1) \Omega - \Sigma_2 =  \pm
                 (1-\Sigma_1) K
\end{equation}
and give the usual dispersion relations \cite{klimov, weldon2,
thoma}. In particular $\Omega_{-} (K)$ has a minimum for non-zero
$K$ and describes the usual plasmino mode.

\section{The three-point funtion and the effective action}

The vertex correction of the three point function,
 $\Gamma^{\mu}(k_1, k_2)$, with two external fermion
 momenta $k_1$ and $k_2$,
can be extracted from
\begin{equation}
\int d^4 x \, d^4 y \, d^4 z \,
     \psi (x) S(x, y) A^{\mu}(y) S(y, z) G(z, x) \bar{\psi} (z)
\end{equation}
using the expressions (\ref{fp}) and (\ref{bp}) and making the same 
hard thermal loop approximations we get
\begin{equation}
e \, \Gamma^{\mu}(k_1, k_2) = m_f^2 \, \int \frac{d\Omega}{4\pi}
\,   \frac{1}{(u\cdot Q)^2}
                  \frac{Q^{\mu} \,(\gamma\cdot Q)}
                       {(k_1 \cdot Q) \, (k_2\cdot Q)}
\label{vertex}
\end{equation}
This can also be written covariantly as:
\begin{equation}
e\, \Gamma^{\mu} (k_1, k_2) = \alpha_1 \, k_1^{\mu} +
                              \alpha_2 \, k_2^{\mu} +
                              \alpha_3 \, u^{\mu}.
\end{equation}
where the expressions $\alpha_1$, $\alpha_2$ and $\alpha_3$ are
covariant, scalar functions  of $k_1$, $k_2$, $u$, and the gamma
matrices, that satisfy
\begin{eqnarray}
& &\alpha_1 \, k_1^2 + \alpha_2 \, (k_1 \cdot k_2) +
                  \alpha_3 \, (u \cdot k_1) = \Sigma (k_2)
\nonumber \\
& & \alpha_1 \, (k_1 \cdot k_2) + \alpha_2 \, k_2^2 +
      \alpha_3 \, (u\cdot k_2) = \Sigma (k_1)
\nonumber \\
& &  \alpha_1 \, (u \cdot k_1) + \alpha_2 (u\cdot k_2) +
     \alpha_3 = J(k_1, k_2)
\end{eqnarray}
with $\Sigma (k)$ given by (\ref{sfermion}) and
\begin{equation}
J(k_1, k_2) = \int d\Omega \,
              \frac{(\gamma \cdot Q)}
        {(u\cdot Q) (k_1 \cdot Q) (k_2 \cdot Q)}
\end{equation}

It is now straightforward to see from these expressions, or
directly from (\ref{fermion}) and (\ref{vertex}) that the Ward
identity
\begin{equation}
e (k_1 + k_2)_{\mu}\, \Gamma^{\mu}(k_1, k_2) =
                     \Sigma(k_1) + \Sigma(k_2)
\label{ward}
\end{equation}
that is essential for maintaining gauge invariance in the hard
thermal loop resummation, is satisfied.

Now I will describe the resulting hard thermal loop effective
action. First, as far as the fermion terms are concerned, it is
easy to see from (\ref{fermion}) that
\begin{equation}
{\cal L}_f = m_f^2 \, \bar{\psi}(x) \,
                   \int \frac{d \Omega}{4 \pi}
                   \frac{1}{(u\cdot Q)^2}
               \frac{\gamma \cdot Q}{(Q\cdot D)}
              \, \psi (x)
\label{lf}
\end{equation}
is the appropriate gauge invariant expression for the effective
fermion Lagrangian in coordinate space, with $D_{\mu} =
\partial_{\mu} + eA_{\mu}$ the covariant derivative, which, upon
expanding the denominator, gives the effective three-point vertex
({\ref{vertex}) as the first correction.

Now, one can apply the usual gauge invariance arguments and power
counting rules, to claim that higher order terms in the expansion
of (\ref{lf}) will give the relevant contributions of the N-point
functions in the hard thermal loop approximation \cite{bpft}.

As far as the photon effective action is concerned, one can
make a similar guess from (\ref{p1}), (\ref{p2}), (\ref{p3}),
and obtain an expression that in the rest frame $u^{\mu}=(1,\vec{0})$
reduces to the known result \cite{bpft}. 
Here I will try
to give a physical interpretation in the nonrelativistic limit.
 In order to do that first we write the gauge potential as
\cite{vpn}
\begin{equation}
A^{\mu}(k) = k^{\mu} \theta + \phi^{\mu} +\beta^{\mu}
\end{equation}
where $\theta$ is a term that can be gauged away to zero,
\begin{equation}
\phi^{\mu} =\left(-\frac{k_i^2}{k^2}\Phi,
             -\frac{k_0 k_i}{k^2} \Phi \right)
\end{equation}
is the electrostatic part and
\begin{equation}
\beta^{\mu}= (0, \beta_i) \,\,\,\,\,\,\,k_i\beta_i =0
\end{equation}
is the transverse part. In terms of these and using the relations
$k_{\mu}\Pi^{\mu\nu}=0$ and $k_i \beta_i =0$ we get from
(\ref{eff})
\begin{eqnarray}
\Gamma = & &m^2\int_k  \Phi(k)\Phi(-k)
                \left[(-1+L)+(2k_i^2\,F_1)(\vec{u}\cdot\vec{k})
                     \right] + \nonumber \\
     &+& \Phi(k)\beta_i(-k) \left[(1-3k_0 F_2)u_i\right]
                 + \nonumber \\
  &+&
  \beta_i(k)\beta_j(-k) \left[(k_0 \,F_2) \delta_{ij} +
             (3k_0 G_2 -F_2)(\vec{u}\cdot\vec{k}) \delta_{ij}
                    \right]
       \label{gi}
\end{eqnarray}

This expression can be written alternatively as
\begin{equation}
\Gamma = \int_k \frac{1}{2} \, \epsilon_{ij}\, E_i(k)E_j(-k)
             \,\,      +\lambda\, \vec{u}\cdot
            \left(\vec{E}(k) \times \vec{B}(-k)\right)
\end{equation}
where the general form of the dielectric tensor that is consistent
with Maxwell's equations is
\begin{eqnarray}
\epsilon_{ij}&=&\left[\epsilon_1 +\epsilon_3(\vec{u}\cdot\vec{k})
                  \right] \frac{k_i k_j}{k_i^2} +
                          \nonumber \\
&+&\left[ \epsilon_2 + \epsilon_4 (\vec{u}\cdot\vec{k})\right]
       \left(\delta_{ij} -\frac{k_i k_j}{k_i^2} \right)
\end{eqnarray}
and the coefficients $\epsilon_{1-4}$ and $\lambda$ are calculated
from (\ref{gi}) to be
\begin{equation}
\epsilon_1=\frac{2m^2}{k_i^2} (1-L)
\end{equation}
\begin{equation}
\epsilon_2 = - \frac{2m^2}{k_0} \,F_2
\end{equation}
\begin{equation}
\epsilon_3 = -4 m^2 \, F_1
\end{equation}
\begin{equation}
\epsilon_4 = \frac{2 m^2}{k_i^2 \,k_0^2}
            \left[ 2k_0 -3k_0\,k_i^2\,G_1
                   -F_2 \,(6k_0^2-k_i^2) \right]
\end{equation}
\begin{equation}
\lambda =\frac{2m^2}{k_i^2} \, (1-3 k_0\,F_2)
\end{equation}

The terms $\epsilon_1$ and $\epsilon_2$ agree with the known
values of the longitudinal and transverse dielectric tensor in the
heat bath rest frame \cite{FTFT}, and $\epsilon_3$ and
$\epsilon_4$ give the first order correction in $\vec{u}$.

As I mentioned in the introduction the additional term,
proportional to $\lambda$, is expected to be there in the case of
electrodynamics of moving dielectrics (\cite{diel} and references
therein). It was derived there using symmetry arguments, and its
value was related to that of the dielectric function. Here, of
course, the physical situation is different, however, a similar
term appears and its value is calculated from the microscopic
theory. In any case this term would be important in the
calculation of physical properties and radiative processes.

\section{Comments}

In this work I described some results on the dependence of the
high temperature effective action on the background plasma
velocity. These and similar results, combined with some
phenomenological model regarding the plasma evolution, would be
presumably relevant in the case of calculations of physical
quantities at finite temperature. I showed that, in the case of
QED, the techniques of hard thermal loop effective actions can be
applied. Similar calculations, with appropriate hard thermal loop
resummations, can be done for the effective action in finite
temperature QCD, with possible applications in the theory of quark
gluon plasma.

\end{document}